\documentclass[prd,onecolumn,preprint,showpacs,groupedaddress,superscriptaddress,nofootinbib,floatfix,preprintnumbers,longbibliography]{revtex4-2}

\usepackage{physics}
\usepackage{xcolor}
\usepackage{graphicx}
\usepackage{subfigure} 
\usepackage{dcolumn}
\usepackage{bm}

\usepackage{comment}

\begin{document}

\title{Above-threshold ionization by polarization-crafted pulses}

\author{Camilo Granados}
\email{camilo.granados@gtiit.edu.cn}
\affiliation{Department of Physics, Guangdong Technion - Israel Institute of Technology, 241 Daxue Road, Shantou, Guangdong, China, 515063}
\affiliation{Technion -- Israel Institute of Technology, Haifa, 32000, Israel}
\affiliation{Guangdong Provincial Key Laboratory of Materials and Technologies for Energy Conversion, Guangdong Technion - Israel Institute of Technology, 241 Daxue Road, Shantou, Guangdong, China, 515063}

\author{Enrique G. Neyra}
    \affiliation{Instituto Balseiro (Universidad Nacional de Cuyo and Comisión Nacional de Energía Atómica) and CONICET CCT Patagonia Norte. Av. Bustillo 9500, Bariloche 8400 (RN), Argentina.}

\author{Lorena Reb\'on}
\affiliation{Instituto de F\'isica de La Plata, CONICET - CCT La Plata, Diag. 119 e/ 63 y 64, La Plata 1900, Buenos Aires, Argentina}
\affiliation{Departamento de F\'isica, Facultad de Ciencias Exactas, Universidad Nacional de La Plata, La Plata, Buenos Aires, Argentina}

\author{Marcelo F. Ciappina}
\email{marcelo.ciappina@gtiit.edu.cn}
\affiliation{Department of Physics, Guangdong Technion - Israel Institute of Technology, 241 Daxue Road, Shantou, Guangdong, China, 515063}
\affiliation{Technion -- Israel Institute of Technology, Haifa, 32000, Israel}
\affiliation{Guangdong Provincial Key Laboratory of Materials and Technologies for Energy Conversion, Guangdong Technion - Israel Institute of Technology, 241 Daxue Road, Shantou, Guangdong, China, 515063}
\date{\today}

\begin{abstract}

Coherent light has revolutionized scientific research, spanning biology, chemistry, and physics. To delve into ultrafast phenomena, the development of high-energy, high-tunable light sources is instrumental. Here, the photo-electric effect is a pivotal tool for dissecting electron correlations and system structures. Particularly, above-threshold ionization (ATI), characterized by simultaneous multi-photon absorption, has been widely explored, both theoretical and experimentally. ATI decouples laser field effects from the structural information carried by photo-electrons, particularly when utilizing ultra-short pulses. In this contribution we study ATI driven by polarization-crafted (PC) pulses, which offer precise control over the electron emission directions, through an accurate change of the polarization state. PC pulses enable the manipulation of electron trajectories, opening up new avenues for understanding and harnessing coherent light. Our work explores how structured light could allow a high degree of control of the emitted photo-electrons.

\end{abstract}

\maketitle
\clearpage
\section{Introduction}

Coherent light has guided, for the last decades, countless advances in different areas of sciences, ranging from biology to chemistry and physics~\cite{Granados1, Feringa1,Misha1,Zewail1,Ciappina}. It is now customary, in many optics laboratories, to explore the frontiers of natural processes at temporal scales close to the electron motion in atoms, molecules and larger structures~\cite{Misha1}. This landscape of research is only possible thanks to the development of ultrashort laser pulses with a duration of the order of attoseconds~\cite{LHuillier, Agostini,krausz}. Advances in this technology towards the creation of table-top setups of higher energy attosecond light sources, which would allow us to investigate nature on unexplored energy and time scales, is one of the cornerstones of attosecond science. 


One of the typical uses of coherent light is the photo-electric effect. Here, both single or multi-photon absorption are key to understand, for example, electron correlations or structural properties of the systems under investigation. The ionization of atoms by the simultaneous absorption of several photons arises as one of the fundamental non-linear effects, together with high-harmonic generation (HHG)~\cite{Lewenstein1,Lewenstein2,Lewenstein3}, where a complete new chapter of the light-matter interaction was necessary to fully describe the seminal experimental observations \cite{Lewenstein1,Kulander1,Lewenstein2,NSDI}. The simultaneous absorption of several photons, known as above-threshold-ionization (ATI), has been investigated during several decades both theoretically and experimentally \cite{Kulander1, ATIExp}. Understanding photo-absorption processes allows us to decouple the influence of the strong laser field in photoelectron dynamics from the structural information that photoelectrons could carry, especially when very short pulses are used~\cite{Milosevic1}. In this sense, ATI is a well established phenomenon to study structural properties of the target systems, either atomic or molecular, since it inherits information on the bound-to-free and re-scattering continuum-continuum electronic transitions. With the advent of carrier-envelope-phase (CEP) stabilization mechanisms, the ATI played a paramount role in demonstrating CEP effects in the observed photo-electron spectra. This was proven experimentally by measuring  correlation between photo-electrons emitted in opposite directions \cite{CEPATI}. Without the CEP stabilization, the effect on the photo-electron spectra normally averages out because the random phase of the electromagnetic field that each pulse carries. In another interesting application, the diffraction pattern generated by re-scattered electrons was used to infer  changes in the molecular bond lengths of oxygen and nitrogen molecules~\cite{diff}. This self-imaging of molecules provides a powerful technique by which the changes of the molecular bond lengths can be followed in real time \cite{diff,diff2}. In addition, with the experimental development of laser light carrying angular momentum \cite{Mirhosseini}, it is now possible to decouple the multi-scattering nature of ATI from the inter-pulses interference effects. New pulse synthesis thus open the door to investigate, understand, and manipulate coherent light with unexplored degrees of freedom~\cite{Ciappina,Oren}.  

From the theoretical point of view, ATI and photo-electron angular distributions can be calculated by solving the time-dependent Schr\"odinger equation (TDSE) \cite{Qprop2,Qprop3,Qprop3_2} or by semi-classical approaches \cite{amini2019symphony}. The main difference between both methods lies on the possibility, when using the latter, of studying the different contributions giving rise to the final photo-electron distributions. However, the solution of the TDSE for all degrees of freedom involved in the problem is often difficult or even impossible to find. By using the strong-field approximation (SFA), Su\'arez et.~al.~\cite{Lewenstein2}, were able to characterise the ionization process in terms of direct and re-scattered mechanisms. This is an step forward to fully decouple the distinct contributions to the ionization process and opens a way to extract structural properties of both atomic and molecular targets. More recently, in Ref. \cite{Skewed} it was demonstrated that, by using amplitude-polarization (AP) attosecond pulses, the emitted electron angular momentum could be controlled. The possibility of tailoring the attosecond pulse polarization has the potential to unlock the investigation of complex physical systems. 

In this work, we investigate the ATI phenomenon driven by polarization-crafted (PC) pulses, for which AP pulses are a special case. The use of PC pulses allows us to control the electron emission direction and presents an approach to control the degree of collimation in the photo-electron emission direction. The complete understanding of the highly focused laser-generated electron beams could be useful for extracting information for both the laser pulse itself and the atomic or molecular target.

The manuscript is organized as follows: In section~\ref{res} we first define PC laser fields and show how to use the time delay, phase and amplitude ratios for crafting their polarization state. In section~\ref{ATIShort} we present atomic-ATI calculations, obtained from the the solution of the three-dimensional (3D)-TDSE, using two short laser fields with variable time delay and phase between them. Here we show how the different electron emission trajectories are defined by the PC laser field. In section~\ref{ATILong} we present the same results as in the previous section, but here using long PC laser pulses and modifying the laser field amplitudes' ratio. Next, in section~\ref{ADistri}, we show the angular dependence for different PC laser fields. Finally, the conclusions are included in section~\ref{sec_conclu}.

\section{Results} \label{res}
We first revise the concept of PC pulses and, then, we show how they can be used to control the electron dynamics. For this purpose, we compute the photo-electron spectra of an hydrogen atom, when interacting with a laser field synthesized from two identical 800-nm ultrafast laser fields with variable relative phase and time delay between them. Our study is divided in two regimes, depending on the temporal duration of the laser pulses, i.e.~we analyse the use of short and long laser pulses. For calculating the photo-electron ATI spectra we used the Q-\textsc{prop} 3D-TDSE solver \cite{Qprop3_2}. 

\subsection{Polarization-crafted pulses}
A PC pulse results from the coherent superposition of two linearly polarized orthogonal electromagnetic fields, i.e., $E(t)=E_{x}(t)\hat{x}+E_{y}(t)\hat{y}$, with no restriction on their relative phase, time delay and relative amplitude. This is in contrast to the AP pulses, for which the relative phase and amplitude of each component is the same. Thus, each of the electric field components, $E_{x}(t)$ and $E_{y}(t)$, can be written as follows: 
\begin{subequations}
\begin{eqnarray}
E_x(t) &=& E_{0,x} \sin^{2}\left(\frac{\omega_0 t}{2 n_c}\right)\sin(\omega_0 t+\phi_x), \label{e1a}\\  
E_y(t) & = & E_{0,y} \sin^{2}\left(\frac{\omega_0 (t-\delta t)}{2 n_c}\right)\sin(\omega_0 (t-\delta t) + \phi_y), \label{e1b}
\end{eqnarray}
\end{subequations}
where $E_{0,x}$ ($E_{0,y}$) is the laser field peak amplitude in the $x$ ($y$)-direction, $\omega_0$ the laser field central frequency, $n_c$ the number of total cycles and $\delta t$ and $\phi_x$ ($\phi_y$) are the time delay and carrier envelope phase (CEP) of the $x$ ($y$) laser field component, respectively. We also define the global phase $\phi=\phi_y-\phi_x$. This quantity is instrumental for the control of the laser field shape, as we will see in the next sections. Notice that by forcing $\phi_x=\phi_y$ and $E_{0,x}=E_{0,y}$ one recovers the well-known AP pulses described, for instance, in Refs.~\cite{Ciappina,Oren}. Here, however, we explore all the possible ways to control the polarization state of the driving laser field. In Ref.~\cite{Ciappina}, the pulse temporal width, used to investigate HHG with Gaussian-shaped AP pulses, was set to $2\Delta t$ and $5\Delta t$ ($\Delta t$ is the full-width at half maximum), corresponding to $\approx$ 6 and 15 total optical cycles, $n_c$. We use a similar set of temporal parameters in the present contribution. 

\subsubsection{Effect of the laser electric field amplitudes' ratio}

\begin{figure}[h!]
\includegraphics[width=1\textwidth, angle = -90]{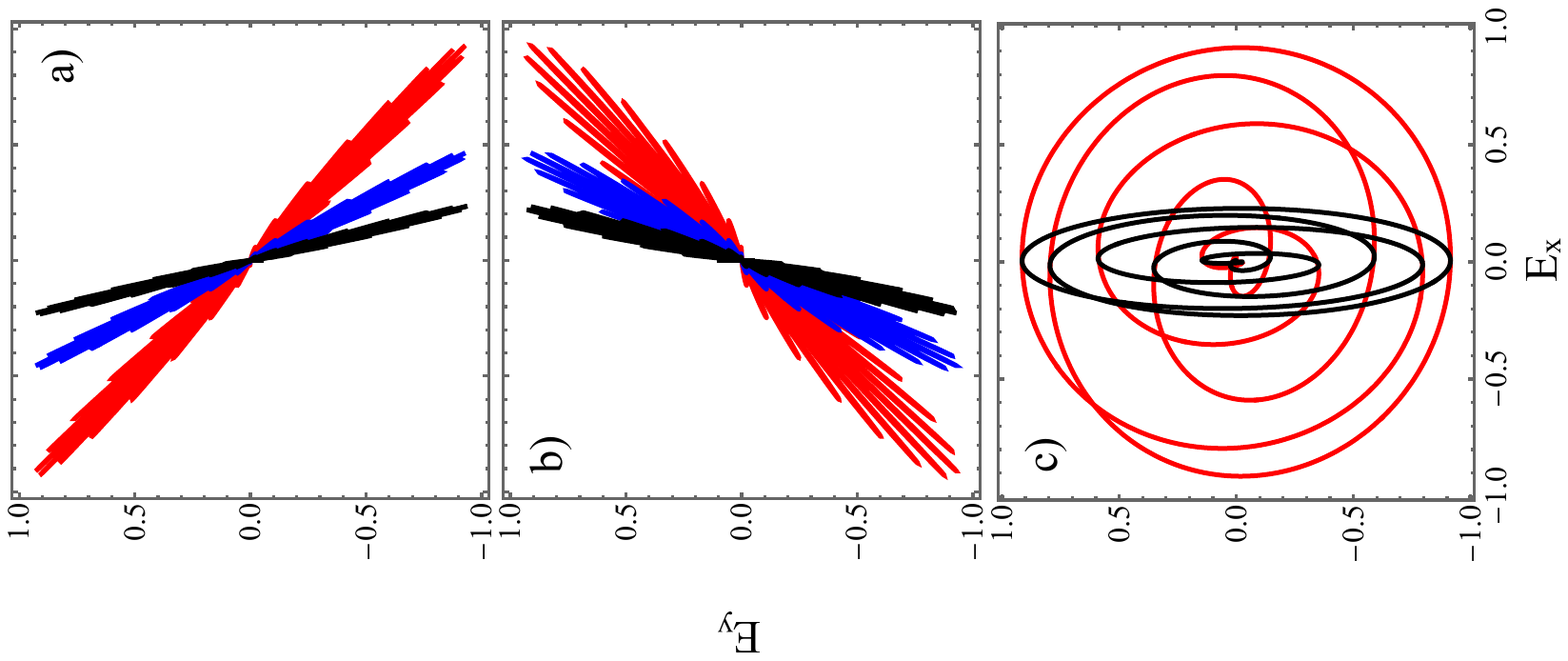}
\caption{\label{fig:fig1} Polarization-crafted (PC) pulses for a laser pulse with $n_c=16$. (a) Lissajous curves for $\delta t=T/2$ and ratios $E_{0,y}/E_{0,x}$ = 1, 1/2 and 1/4 in red, blue and black, respectively. (b) same as panel (a) but with $\delta t=T$.
(c) Lissajous curves for $E_{0,y}/E_{0,x}$ = 1 and 1/4 with $\delta t=T/4$. In all cases, it is $\phi=0$.}   
\end{figure}

By changing the ratio between the laser field amplitude components in Eqs.~(\ref{e1a}) and~(\ref{e1b}), for a fixed time delay $\delta t$, it is possible to change the angle of the total laser field with respect to the plane $E_y=0$, i.e.~we obtain a linear-like Lissajous curve. Furthermore, for certain values of $\delta t$ (see next section), we get a circular-like Lissajous curve. Here, the change in the ratio shrinks the total laser field in the direction of the component with smaller amplitude. We show these general features in Figs.~\ref{fig:fig1}(a) and \ref{fig:fig1}(b),
where we depict the Lissajous curves for 3 different laser field amplitudes' ratios $E_{0,y}/E_{0,x} = $ 1, 1/2 and 1/4 (red, blue and black curves in Figs.~\ref{fig:fig1}(a) and (b), respectively), and two different time delays $\delta t = T/2$ in Fig.~\ref{fig:fig1}(a) ($T=2\pi/\omega_0$ is the optical period) and $\delta t = T$ in Fig.~\ref{fig:fig1}(b), keeping the global phase at the value $\phi=0$. Finally, in Fig.~\ref{fig:fig1}(c) we show the resulting Lissajous figures for a time delay corresponding to $\delta t = T/4$ and two different amplitudes' ratios, $E_{0,y}/E_{0,x} = $ 1 and 1/4, in red and black, respectively. 
Experimentally, changing the ratio between the amplitudes of the laser field components corresponds to change the angle on the initial incident laser field before entering the birefringent media used to set their relative time delay~\cite{Karras2015,Ciappina}. This initial incident angle corresponds, in our current description, to the global angle $\theta$. For $E_{0,y}/E_{0,x} = $ 1, $\theta$ takes the values $-\pi/4$ or $\pi/4$, for $\delta t = T/2$ and $T$, respectively. When the ratio is $E_{0,y}/E_{0,x} =1/2$ ($E_{0,y}/E_{0,x} =1/4$), $\theta$ takes the value $0.35\pi$ ($0.45\pi$). In the next section we show how to analytically calculate $\theta$.

\subsubsection{Effect of the time delay}
\begin{figure}[h!]
\includegraphics[width=1\textwidth, angle = 0]{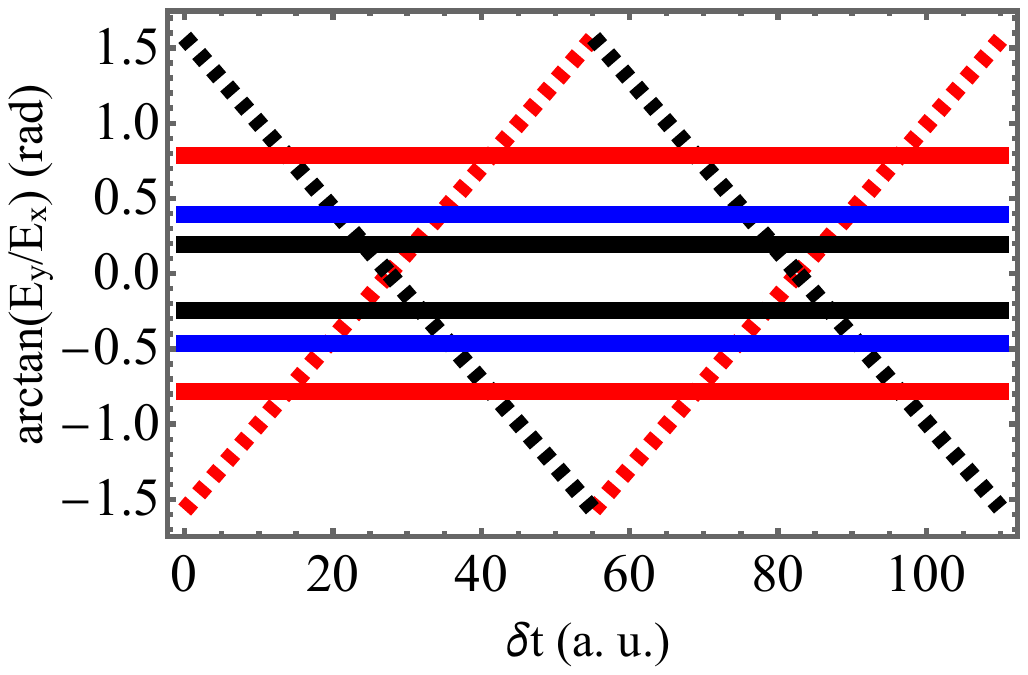}
\vspace{-3cm}
\caption{\label{fig:fig1.1} Temporal evolution of the global angle, $\theta$, calculated from Eq.~(\ref{eq2}). The red, blue and black solid curves are the $\theta$ values for the parameters used in Fig.~\ref{fig:fig1}(a). In the same plot, the dashed red and dashed black curves are for the case of $E_{0,y}/E_{0,x}=1$ and $\delta t=T/4$ and $\delta t=3T/4$, respectively.} 
\end{figure}

By changing the time delay $\delta t$ in Eq.~(\ref{e1b}), it is also possible to modify the shape of the Lissajous curve formed by the two orthogonal laser fields. For example, if the delay is set to $\delta t=nT/2$, with $n$ an integer, the Lissajous curves have a distinct line-like shape. However, if  $\delta t=(2n-1)T/4$ the Lissajous curves change to a circular-like shape. Thus, the latter value of $\delta t$ defines a circular polarized laser field with time-dependent ellipticity. This time-dependent ellipticity steers the electron trajectories in such a way that the electron re-scattering with the parent ion is avoided.
For a fixed laser electric field amplitudes' ratio, $E_{0,y}/E_{0,x}$ it is possible to obtain the evolution of the global angle, $\theta$, as a function of  the time delay, $\delta t$. Here $\theta$ is considered as the polarization angle with respect to the plane $E_x=0$. To calculate $\theta$ we use the expression $\tan(\theta)=A\sin(\omega (t-\delta t))/\sin(\omega t)$, where we have defined $A=E_{0,y}/E_{0,x}$. Then,
\begin{equation}
    \theta=\arctan\Bigg(A\frac{\sin(\omega(t-\delta t))}{\sin(\omega t)}\Bigg) =\begin{cases}
	 - \arctan\Big(A\cot(\omega t)\Big), & \text{for }  \delta t =T/4\\		
   -A\pi/4, & \text{for }  \delta t =T/2\\        
           \arctan\Big(A\cot(\omega t)\Big), & \text{for }  \delta t =3T/4\\
      A\pi/4, & \text{for } \delta t =T.
		 \end{cases}
   \label{eq2}
\end{equation}
The different cases of Eq.~(\ref{eq2}) are presented in Fig.~\ref{fig:fig1.1}. Here, we depict  in red, blue and black, the results of Eq.~(\ref{eq2}) with a time delay of $\delta t = T/2$ and for $E_{0,y}/E_{0,x} = 1, 1/2 $ and 1/4, respectively. The dashed red and dashed black lines show the case for $E_{0,y}/E_{0,x}= 1$ and $\delta t = T/4$ and $\delta t = 3T/4$, respectively. We can see that (i) $\tan(\theta)$ is constant during one full laser cycle for $\delta t = n T/2$, and (ii) $\tan(\theta)$ varies with time for  $\delta t = (n+1) T/4$. 

From Eq.~(\ref{eq2}) we can define an AP laser field with constant polarization, e.g.~setting $\theta = \pi/4$. Likewise, Eq.~(\ref{eq2}) includes AP laser fields with variable polarization as well, when $\theta =\arctan\big(\cot(\omega t)\big)$. Notice that AP pulses are only possible if $\phi_x=\phi_y=0$ and $A=1$. From PC pulses, changes in time delays, $\delta t$, and amplitude ratios, $E_{0,y}/E_{0,x}$, allow us to tailor the global angle $\theta$ and polarization state of the AP laser fields. Thus, we could have a more precise control of the electron trajectories. 

\begin{figure}[h!]
\includegraphics[width=0.8\textwidth, angle = -90]{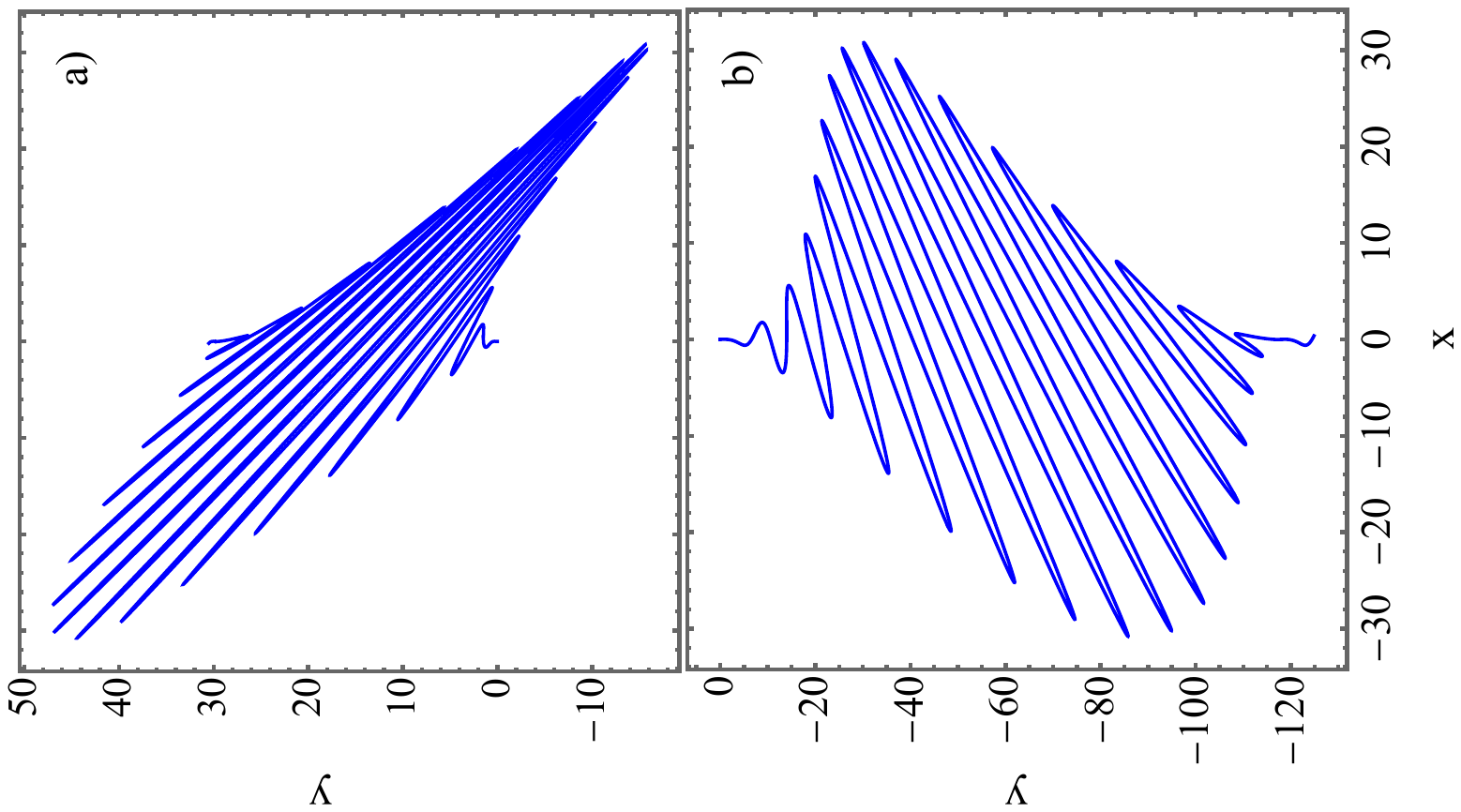}
\caption{\label{fig:fig1.3} Electron trajectories for (a) $\delta t=T/2$ and (b) $\delta t=T$ (red curves in Fig.~\ref{fig:fig1}(a)) calculated by integrating the two-dimensional classical equations of motion of an electron in the laser electric fields given by Eqs.~(\ref{e1a}) and (\ref{e1b}).} 
\end{figure}

To show this possibility, we calculate the resulting electron trajectories by solving the two-dimensional classical equations of motion for one free electron in the laser electric fields defined by Eqs.~(\ref{e1a}) and~(\ref{e1b}). Here, we set the time delays between the pulses as $\delta t = T/2$ and $\delta t = T$. The results are presented in Figs.~\ref{fig:fig1.3}(a) and \ref{fig:fig1.3}(b), respectively. Our calculations show good agreement with the parametric plot of the fields presented in Figs.~\ref{fig:fig1}(a) and~\ref{fig:fig1}(b). It is important to mention that the classical trajectories show a larger angular dispersion as the delay between the pulses increases. This will have a clear impact on the ATI spectra, as we show in the next section.

\subsubsection{Angular dispersion of the electron trajectories}
The angle between successive laser field oscillations is useful to investigate the angular dispersion of the electron trajectories. For simplicity we call these oscillations ``petals". To calculate such an angle, we compute the ratio between the laser electric field components envelopes. By expanding the ratio and taking only the linear terms in $t$ and $\delta t$ we arrive to: 

\begin{equation}
    \alpha=\arctan\Bigg(A \frac{\sin^2(\omega(t-\delta t)/(2 n_c))}{\sin^2(\omega t/(2 n_c))}\Bigg) -A\frac{\pi}{4}\approx  \arctan\Bigg(\frac{2}{3}A\frac{\omega^2t}{n_c^2}\delta t \Bigg) - A\frac{\pi}{4}+O(\delta t^2).
   \label{eq3}
\end{equation}
For a laser field with $n_c =6$ cycles, $\omega=$ 800~nm,  $\delta t = T/2$ and $t=T/2$, the angle between petals is $\alpha\approx 10^{\text{o}}$. For the same laser parameters, but with $n_c =15$ cycles, Eq.~(\ref{eq3}) predicts a smaller angle, $\alpha\approx 1^{\text{o}}$. Thus, by increasing the number of cycles we can control the collimation of the electron trajectories when PC laser fields are used. 
Comparing Eq.~(\ref{eq3}) with the results presented in \cite{Ciappina}, we see a smaller increase in the angle as the time delay increases. However, the functionality of Eq.~(\ref{eq3}) is more complete, considering we have kept the ratio between the orthogonal laser field amplitudes. This ratio plays a fundamental role on the angular dispersion of the Lissajous curves. 
Overall, from Eq.~(\ref{eq3}), we can conclude that the angle between successive petals increases with the time delay, making the angular dispersion of the petals larger. Contrarily, for a fixed time delay, by increasing the number of cycles, the electron trajectories become more collimated around the global angle, $\theta$.  Likewise, we can use the ratio between the laser fields amplitudes to change the angular dispersion of the electron trajectories and the angle $\alpha$, as well as the global angle $\theta$ of the Lissajous curve, as mentioned previously. This means that we can vary the amplitudes ratio to select the electron emission angle, as already shown in Fig.~\ref{fig:fig1}(a). This is consistent with the fact that, reducing the amplitude of one of the laser fields components, the excursion of the electron in such a direction gets reduced as well. 
It is important to remark that the calculation of the angular dispersion dependence with the number of cycles and the time delay between the laser electric field components is only valid for time delays that are multiple integers of $T/2$. For time delays that are multiple integers of $T/4$, a constant dispersion angle has not meaning since it changes with time, as shown by Eq.~(\ref{eq2}).

\subsubsection{Effect of the global phase}

\begin{figure}[h!]
\includegraphics[width=0.9\textwidth, angle =0]{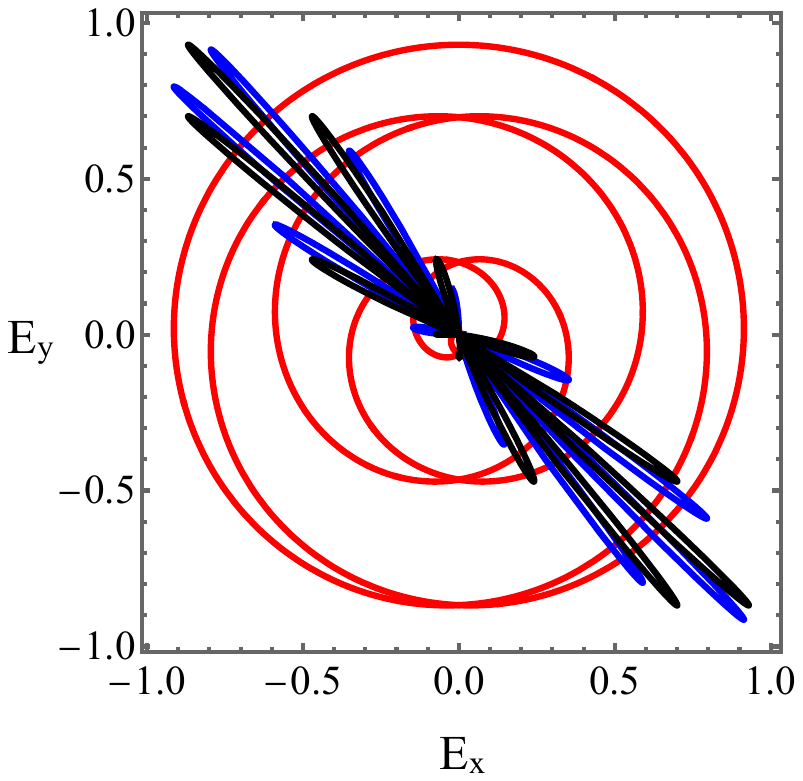}
\caption{\label{fig:fig21} Parametric plots for different values of the laser field global phase, $\phi$. In red and blue  we plot the field's parametric plot with $\phi=\pi/2$ and$\phi=0$ rad. The time delay for both cases is the same, $\delta t = 0$ and T/2. Notice that for the case of AP pulses, a global phase change of $\pi$ will only change the position of the petals and not the shape of the total laser field. This is shown by the black curve for which we used $\phi=0$ but with each laser field phase given by $\phi=\pi/2$ and a time delay between the pulses of $\delta t =T/2$. }
\end{figure} 

Until now, we have demonstrated control over electron trajectories by tuning the laser electric field components amplitudes ratio, $E_{0,y}/E_{0,x}$ and its relative time delay, $\delta t$. In addition, by modifying the CEPs $\phi_x$ and $\phi_y$ of each laser electric field components, we can change the Lissajous curve from a circular-like to a line-like shape. Using the global phase $\phi$ we have then, three different methods to control the electron dynamics. The difference is, however, that by changing the Lissajous curve from the time delay $\delta t$, the angular dispersion of the electron trajectories changes. The same modification in the Lissajous curves is obtained by varying the global phase $\phi$, with the difference that this method leaves the dispersion angle unchanged. 

We show in Fig.~\ref{fig:fig21} the parametric plots of the laser field components for a constant time delay $\delta t = T/2$, and two different global phases, namely $\phi = 0$ (red solid line) and $\phi = \pi/2$ (blue solid line). Here, the most remarkable thing is that the Lissajous curve has a well defined shape for a fixed time delay, $\delta t$ and a fixed global phase $\phi$ (solid black line). Experimentally, this opens the possibility of characterizing the CEP: for example, by measuring the photo-electrons produced by ultrashort PC laser fields one could extract variations in the CEP from pulse to pulse if the time delay between the pulses is fixed. This is because the time delay, $\delta t$, together with the amplitude ratio of the laser filed, $E_{0,y}/E_{0,x}$, and the global phase $\phi$ define a unique Lissajous curve, which consequently gives rise to a unique photo-electron ATI spectra. This technique, however, would be limited to the degree of time delay stabilization. More importantly, the PC laser fields could provide a novel way to characterize the CEP for pulses with many cycles, where the direct CEP measurement is difficult. 

\section{ATI with short PC fields} \label{ATIShort}

In the previous sections, we demonstrated how the manipulation of the time delay, $\delta t$, relative phase $\phi$, and field's components amplitude ratio $E_{0,y}/E_{0,x}$, allowed us to control the electron trajectories. Additionally, we illustrated that specific time delays can result in a laser pulse exhibiting time-dependent ellipticity. This occurs when  $\delta t =(2n-1)T/4$. Furthermore, the time delay between the laser electric field components determines whether the electron undergoes multiple rescatterings with the parent ion or not. For example, a delay of $\delta t = T/4$ leads to the total absence of rescattering. Conversely, for $\delta t = T/2$, there are multiple rescattering events as the electron's trajectory intersects the ion position repeatedly. It is then instrumental to distinguish between these features to understand if the observed photo-electron spectra is formed by multiple rescattering events and/or by intra-cycle interferences~\cite{Qprop3_2}. This distinction is particularly important for molecular structure characterization~\cite{diff}.

\begin{figure}[h!]
\includegraphics[width=1\textwidth]{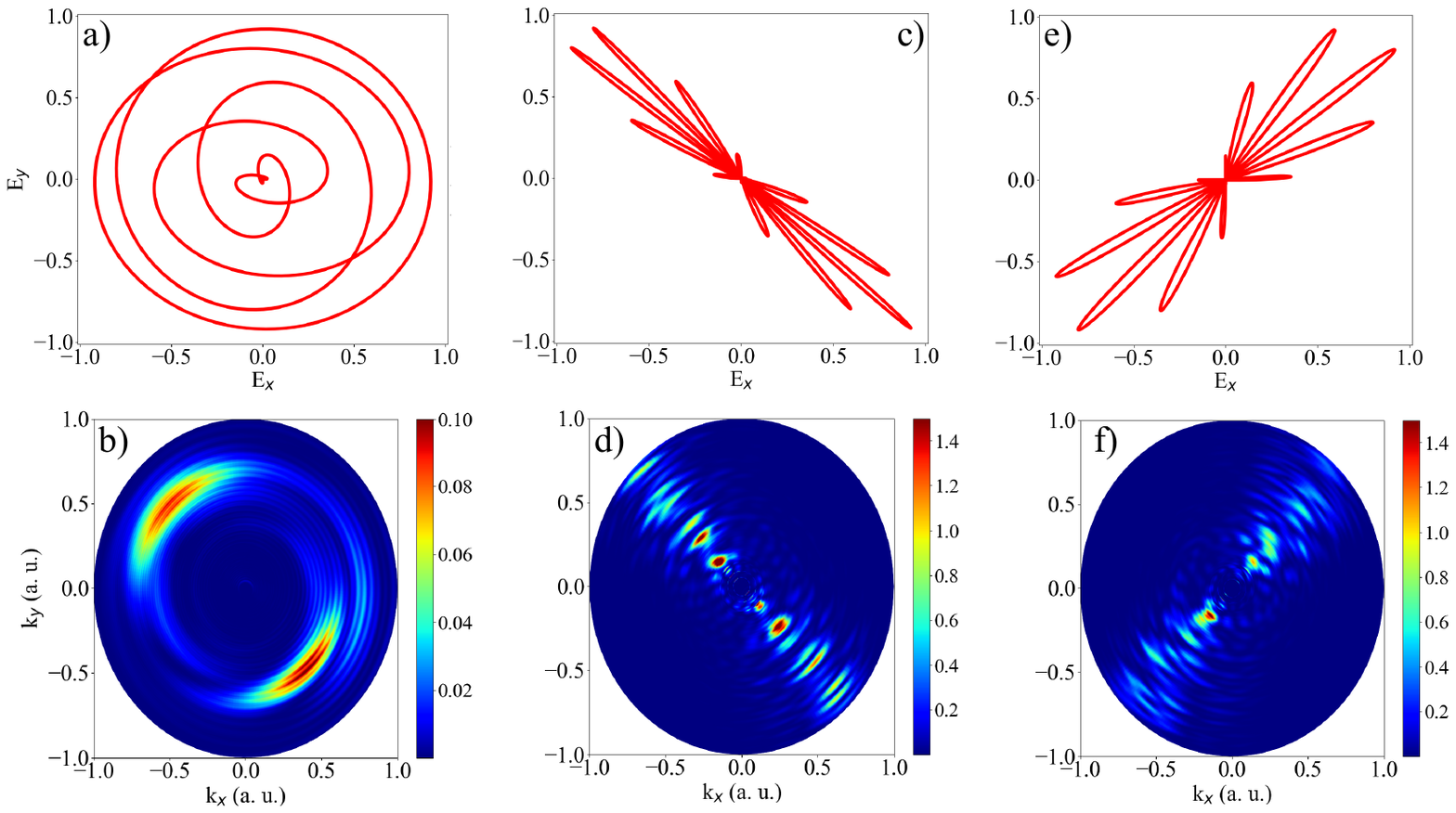}
\caption{\label{fig:fig2} Parametric plots of the laser field components and generated photo-electron spectra for the case of short PC pulses. The relative time delay between the laser fields components is $\delta t = T/4$ for (a) and (b), $\delta t = T/2$ for (c) and (d) and $\delta t = T$ for (e) and (f). It is clear from (d) and (f) that the angular dispersion of the emitted electrons increases with the time delay.}
\end{figure} 

In order to study how the different PC-pulse configurations tailor the photo-electron spectra, we drive an hydrogen atom with short and long laser PC pulses and compute the 2D photo-electron momentum distributions, i.e.~the probability to emit electrons parallel and perpendicular to the laser electric field polarization, using the 3D-TDSE~\cite{Qprop3_2}. For the case of short PC pulses, we set the number of cycles $n_c=6$ (total duration $\approx 16$~fs) and the laser field central wavelength $\lambda =$ 800~nm ($\omega=0.057$ a.u.). Furthermore, the laser field amplitudes of each component are set to $E_{0,x}=0.0534$ a.u. and $E_{0,y}=0.0534$ a.u.~and their CEPs as $\phi_x =\phi_y= 0$. The calculated ATI 2D photo-electron momentum distributions and laser electric field components are presented in Fig.~\ref{fig:fig2}. Figure~\ref{fig:fig2}(a) shows the parametric plot of the laser field components with a time delay $\delta t = T/4$. The resulting ATI spectrum is depicted in Fig.~\ref{fig:fig2}(b). Here, the shape of the 2D photo-electron momentum distribution is typical for atoms driven by circularly polarized fields~\cite{Milosevic}. In Figs.~\ref{fig:fig2}(c) and (d) we show the corresponding parametric plot of the laser fields and the resulting 2D photo-electron momentum distribution, respectively, for the case of a time delay $\delta t = T/2$.  Finally, Figs.~\ref{fig:fig2}(e) and (f) present the parametric plot of the laser fields and the resulting 2D photo-electron momentum distribution, respectively, for $\delta t = T$. Several characteristics predicted by Eq.~(\ref{eq3}) are clear in the computed ATI spectra, namely 1) The electron trajectories closely follow the polarization direction of the PC pulses, 2) it is evident in all the plots (Figs.~\ref{fig:fig2}(c) to (f)) that, for the specific time delay used and an amplitudes ratio $E_{0,x}/E_{0,y}=1$, the global angle is $\pm \pi/4$ and 3) the electron trajectories angular dispersion angle, $\alpha$, increases with the time delay between the laser electric field components. This results in a larger angular dispersion in the 2D photo-electron momentum distribution, as shown in Figs.~\ref{fig:fig2}(d) and (f).

\section{ATI with long PC fields}\label{ATILong}

\begin{figure}[h!]
\includegraphics[width=1\textwidth]{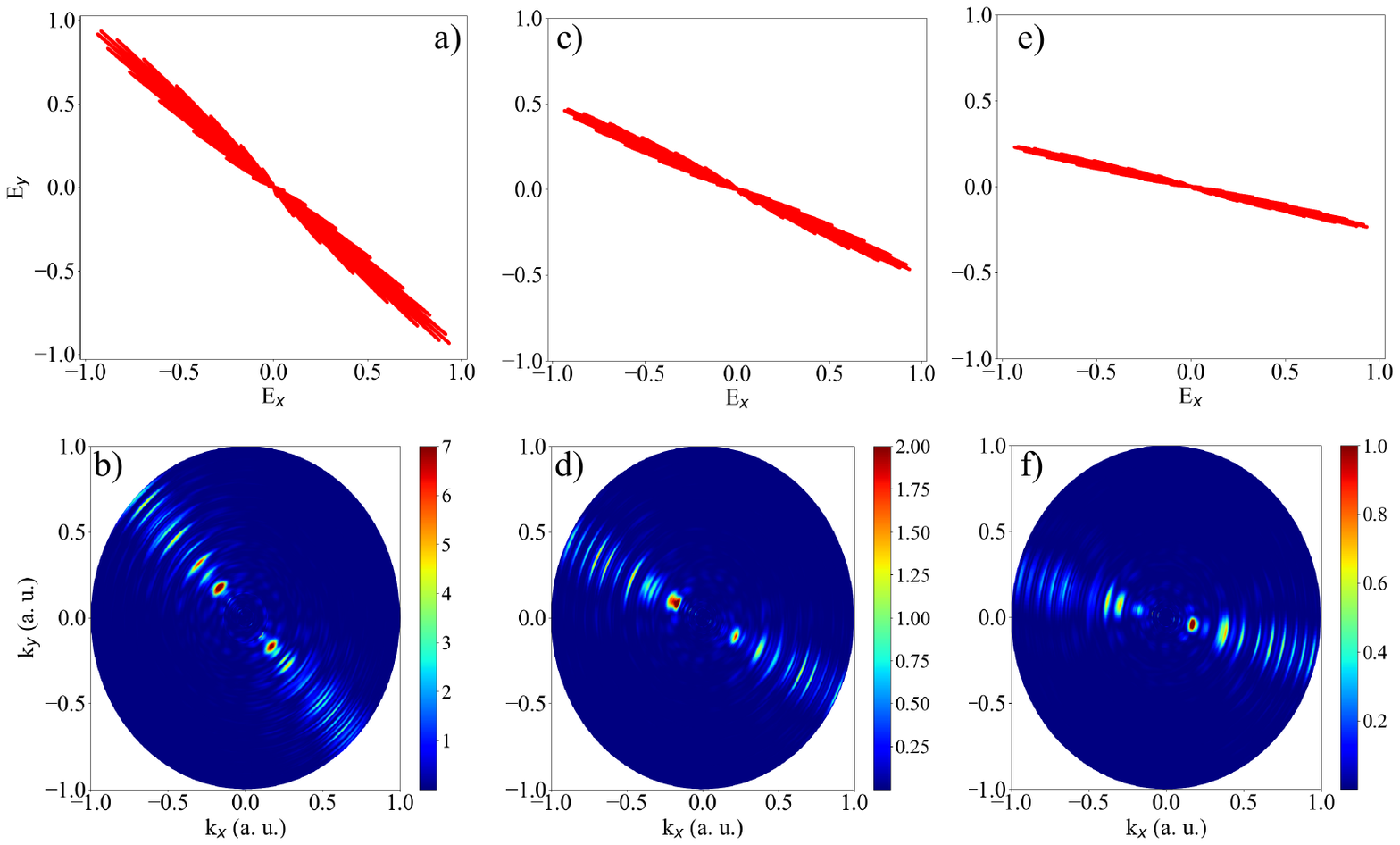}
\caption{\label{fig:fig3} Parametric plots of the laser field components and generated 2D photoelectron spectra for the case of long PC pulses. Here $n_c$ = 16 and the time delay $\delta t = T/2$. (a) and (b) $E_{0,y}/E_{0,x}=1$, (c) and (d) $E_{0,y}/E_{0,x}=2$ and (e) and (f) and $E_{0,y}/E_{0,x}=4$.}
\end{figure}

We also investigate the 2D photo-electron momentum distributions for long PC pulses. This case is of great interest since it can provide a way to characterize the CEP for long laser pulses. Here, we use $n_c =$ 16 cycles ($\approx 43$~fs) and $\phi_x=\phi_y=0$. We show, in Fig. \ref{fig:fig3}, the parametric plots for the laser field with different laser field amplitude ratios, namely (a) $E_{0,y}/E_{0,x}=1$, (c) $E_{0,y}/E_{0,x}=2$ and (e) $E_{0,y}/E_{0,x}=4$. The resulting 2D photo-electron momentum distributions are presented in panels (b), (d) and (f), respectively. In all these cases we have set $\delta t = T/2$. Once again, as predicted by Eq.~(\ref{eq3}), we can see the control of the photoelectrons global angle by changing the laser fields amplitudes ratio. Note also the changes in the amplitudes in the differential ionization probability for different global angles (see the color bars). As mentioned before, experimentally, this scenario corresponds to change the initial angle of the fundamental field. 

\section{Angular distribution}\label{ADistri}

As shown in the previous section, the photo-emission spectra generated with PC pulses (see Figs.~\ref{fig:fig2} and~\ref{fig:fig3}) follows closely the laser electric field. It is, however, important to characterize the angular distribution of the differential ionization probability,  in order to quantify up to what order we can control the angle $\theta$ and, in the case of AP pulses, $\alpha$.  We compare this latter case with the resulting photo-electron spectra from a laser field with variable polarization (corresponding to $\delta t=(2n-1)T/4)$). 

\begin{figure}[h!]
\includegraphics[width=1\textwidth]{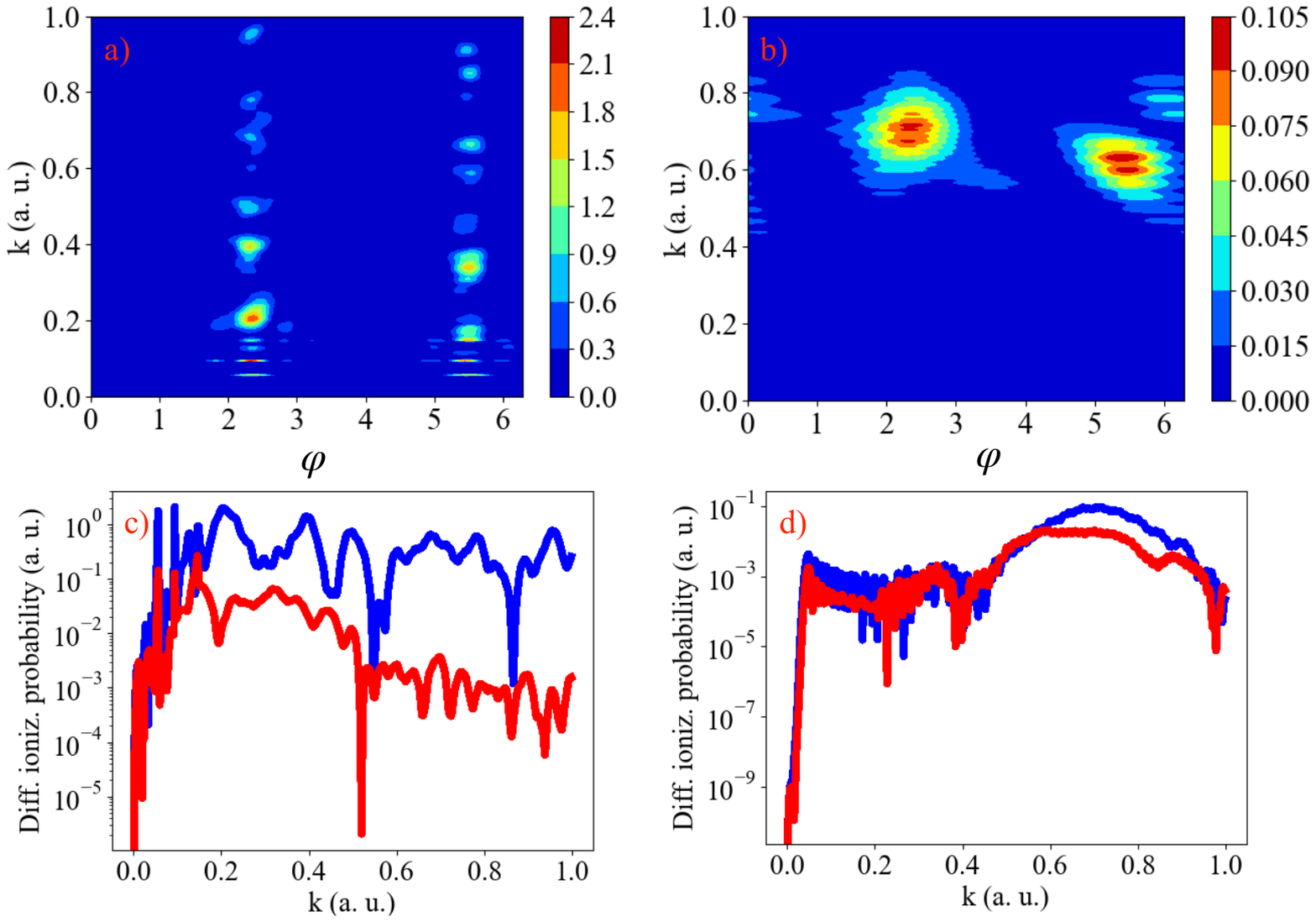}
\caption{\label{fig:fig4} Angular dependency of the ATI spectra.
Angular dispersion (a)-(b) and differential ionization probability (c)-(d) extracted from the ATI spectra when a) $\delta t = T/2$ and b) $\delta t = T/4$. The lower plots show the differential ionization probability for the angles $\phi = 2.3$ (blue line) and 3~rad (red line). The ratio between the maximum values for the red and blue curves in (a) is $\approx 5$ and in (b) is $\approx$ 22.}
\end{figure} 

To understand the differences in the angular dispersion between the particular cases of AP and polarization dependent pulses, in Fig. \ref{fig:fig4} we show contour plots with the differential ionization probability as a function of the global angle, $\theta$, and the photo-electron momentum, $k$. The ATI spectra were computed using different values for time delay, namely (a) $\delta t =T/2$ and (b) $\delta t =T/4$. The photo-electron distribution for $\delta t =T/2$ shows a collimated distribution around the angles $3\pi/4$ and $7\pi/4$. Likewise, for $\delta t =T/4$ the electron distribution maxima are larger around the same angles, but they exhibit a collimation in the electron momentum as well (see the ``islands" around to $k\approx0.6-0.8$ a. u.). In the same plots, it can be seen that the differential ionization probability is larger for $\delta t =T/2$ (see color bar scale in Figs.~\ref{fig:fig4} (a) and (b)). As discussed in Ref.~\cite{Qprop3}, fields with variable polarization, as the one used to obtain Fig.~\ref{fig:fig4}(b), drive the electron away from the parent ion, avoiding re-scattering. There exists, however, cases where multiple re-scattering events are present (see Fig.~\ref{fig:fig4}(a)).  

In Figs.~\ref{fig:fig4}(c) and (d) we show, in blue, the differential ionization probability as a function of the electron momentum for the same cases shown in (a) and (b) and the angle, $\phi = 2.3$ rad. The red curves show the differential ionization probability for $\phi=$ 3 rad. While the differential ionization probability in Fig.~\ref{fig:fig4} (c) shows a large decrease when  $\phi\rightarrow 2\pi$, in Fig.~\ref{fig:fig4}(d) the change in the differential ionization probability for the same global angle values is smaller. To quantify this, we calculate the ratio between the differential ionization probability maximum values for the two different angles. In the case of $\delta t =T/2$ the ratio corresponds to $\approx$ 22 while for $\delta t =T/4$ the ratio is $\approx$ 5. This clearly proves that it is possible to control the photoelectron angular distribution with different PC pulses. 

\begin{figure}[h!]
\includegraphics[width=1\textwidth]{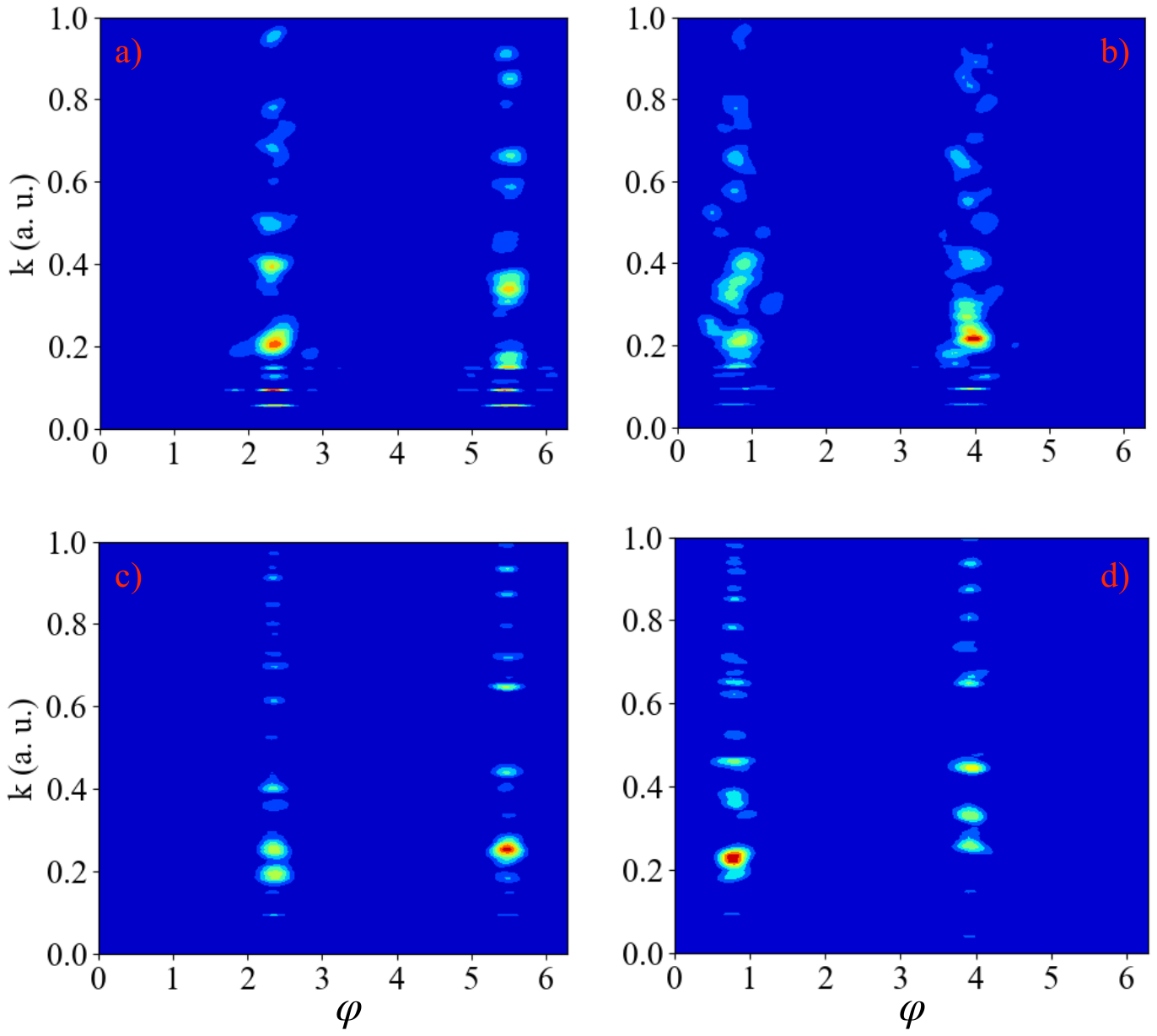}
\caption{\label{fig:fig5} Photoelectrons angular dispersion. In panels (a) and (b) the case for short pulses, while the case for long pulses it is shown in panels (c) and (d). Here, we have used two different delays, i.e., $\delta t = T/2$ for (a) and (b) and $\delta t = T$ for (d) and (d).}
\end{figure}

We now compare the change of the photoelectron angular distribution as a function of the time delay and number of cycles. In Fig.~\ref{fig:fig5} we show the resulting photoelectron distributions. In (a) and (b) we use short PC pulses and in (c) and (d) long PC pulses. In all the cases we set $\delta t =T/2$. As described by Eq.~(\ref{eq3}), by increasing the number of cycles, we observe a smaller angular dispersion (compare (a) and (c) and (b) and (d)). Also, let us notice that the effect of the time delay on the angular dispersion is larger for short pulses than for the long ones. This is also observed in the 2D photoelectron momentum distributions shown in Figs.~\ref{fig:fig2} and~\ref{fig:fig3}. 

\begin{figure}[h!]
\includegraphics[width=1\textwidth]{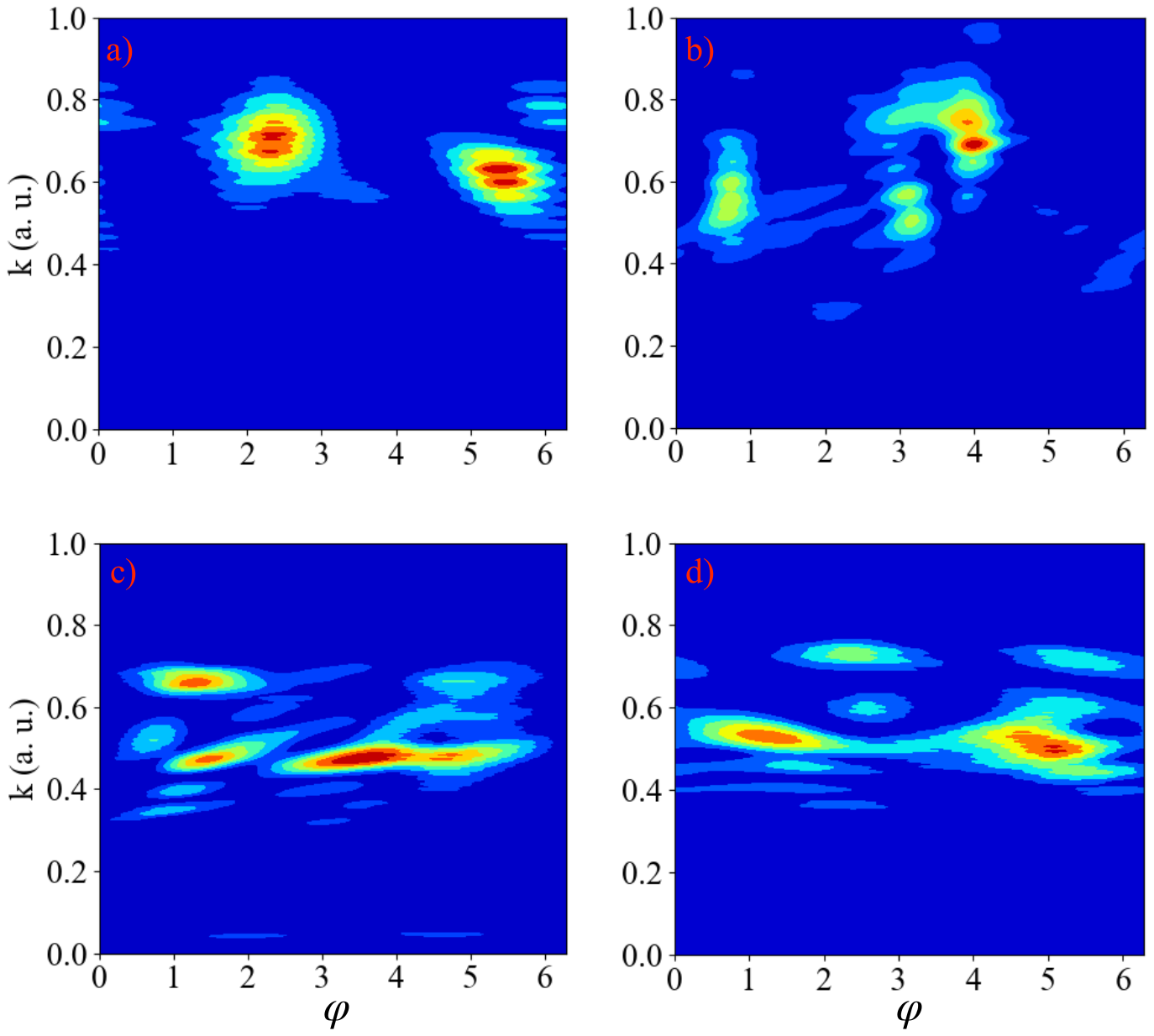}
\caption{\label{fig:fig6} Photo-electrons angular dispersion for $\delta t = (2n-1)$T/4. In a) and b) short pulses and c) and d) long pulses. Here, we have used two different delays, i.e., $\delta t = T/4$ (a) and (c) and $\delta t = 3T/4$ (b) and (d). }
\end{figure} 

The plots in Fig.~\ref{fig:fig5}, contrast with the photo-electron angular distribution for the laser fields with variable polarization, as seen in Fig.~\ref{fig:fig6}. While the angular distribution is collimated around the global angle (given by the laser field components' amplitude ratio) for the AP pulses, when variable polarization pulses are used, the photo-electron signal is distributed around larger angles. This angle increases with the number of cycles and the time delay. In Figs.~\ref{fig:fig6}(a) and (b) we plot the photo-electron angular distribution for $\delta t =T/4$ and $3T/4$ and $n_c=6$ cycles, meanwhile in Figs.~\ref{fig:fig6}(c) and (d)  we set $\delta t =T/4$ and $3T/4$ and $n_c=15$ cycles. It is clear from these figures that by increasing number of pulses both the angular and momentum dispersion increase.  

\section{Conclusions}\label{sec_conclu}

We demonstrated control over the electron emission trajectories by solving the 3D-TDSE for above-threshold ionization (ATI) with short and long polarization-crafted (PC) laser fields. The uniqueness of these laser fields lies in the possibility to finely control the laser field polarization using three different parameters: the relative phase, the time delay, and the laser field's amplitude ratio between the orthogonal components. These various control parameters are crucial for altering the global angle of the electron trajectories and the angle between the oscillations of the total laser field. Since the electron trajectories are well characterized by these parameters, additional information is available in the resulting ATI spectra. By fixing the time delay and the amplitude ratio, one can characterize the CEP of the pulse independently, whether the laser fields are short or long. The latter case is particularly significant, as retrieving long pulse CEP information is exceptionally challenging. Similarly, by fixing any two parameters, one can extract clear information about the remaining one.

The level of characterization and control achieved with the ATI spectra is fundamental for fully understanding the structural information that ATI could provide in more complex target experiments. Furthermore, while the experimental realization of different polarization-crafted laser fields may be challenging, it is within reach for today's ultrafast laser laboratories
 
\section*{Acknowledgements}

The present work is supported by the National Key Research and Development Program of China (Grant No.~2023YFA1407100) and Guangdong Province Science and Technology Major Project (Future functional materials under extreme conditions - 2021B0301030005). M. F. C. and C. G. acknowledge financial support from the Guangdong Natural Science Foundation (General Program project No. 2023A1515010871). E. G. N. and L. R.~acknowledge to Consejo Nacional de Investigaciones Cient\'ificas y T\'ecnicas (CONICET). 

\bibliography{main}

\end{document}


\title{
Twisting attosecond pulse trains by amplitude-polarization IR pulses}
\section*{Supplementary information}

\subsection{Peak amplitude of the
synthesized AP field}\label{suppl_peak}

According to the definition of amplitude-polarized (AP) electric fields (see main text) and assuming a Gaussian-shaped envelope, the explicit expressions for the components of $\bm{E}(t)$, when its polarization direction is at an angle of $45^{\circ}$, are:
\begin{subequations}
\begin{eqnarray}
E_x(t) &=&  E_0\;e^{-2\ln(2)\left(\frac{t+\tau/2}{\Delta t}\right)^2}\;\mathrm{cos}(\omega_0(t + \tau/2 ) + \phi) \label{e1a}\\  
E_y(t) & = & E_0\;e^{-2\ln(2)\left(\frac{t - \tau/2}{\Delta t}\right)^2}\;\mathrm{cos}(\omega_0(t - \tau/2 ) + \phi). \label{e1b}
\end{eqnarray}
\end{subequations}

In the central temporal region, i.e.~around $t=0$, the peak field amplitude ($\left | \bm{E}(t) \right|_{\mathrm{max}} = \left( \sqrt{E_x(t)^2+E_y(t)^2}\right)_{\mathrm{max}}$) for such an AP pulse results
\begin{eqnarray}
    \left | \bm{E}(0) \right|_{\mathrm{max}} = \sqrt{2}E_0 e^{-2\ln(2)\left(\frac{\tau/2}{\Delta t}\right)^2}.
\end{eqnarray} 

Therefore, when $\tau=\Delta t$, we have
\begin{eqnarray}
\left | \bm{E}(0) \right |_{\mathrm{max}} = \sqrt{2}E_0 e^{-\frac{1}{2}\ln(2)} = E_0.    
\end{eqnarray}

\subsection{Lissajous curves of the IR--AP pulse: Angle between two adjacent petals}\label{suppl_angles}

 In this section, we 
 describe how 
 the angle $\alpha$ between two adjacent ``petals" in the Lissajous curves of the driving AP field, varies as a function of the temporal delay $\tau$. The importance of this angle lies in the fact that it represents the angular difference between the polarization directions of two adjacent attosecond pulses in the twisted APT, near the center of the temporal region.

 \begin{figure}[h!]
 \includegraphics[width=0.5\textwidth]{Fig1SM.eps}
 \caption*{\textbf{Supplementary Fig. 1: Lissajous curve of the electric field of the AP pulse with an FWHM $\Delta t=2$ opt. cycles.} The blue and black dashed lines indicate the angular position of the central petal and 
 its first adjacent petal, respectively.}\label{fig:Fig1Sb}
 \end{figure}

 Supplementary Figure 1 shows the Lissajous curve of the electric field of an AP pulse whose temporal FWHM is $\Delta t=2$ opt. cycles. In this figure, 
 $\alpha$ corresponds to the angular separation 
 between the black and blue dashed lines. The blue dashed line intersects the position of the maximum 
 field amplitude at 45$^{\circ}$ (global phase $\phi=0$) 
 while the black dashed line intersects the position of the adjacent maximum. 
 Mathematically, we can find 
 the angle $\alpha$ after evaluating the field ${\bm E}(t)$ at $t=1/2$ opt. cycle, which corresponds to the temporal position of the petal indicated 
 by the black dashed line in the figure, 
 and taking the angular distance to the angle $\frac{\pi}{4}$, which gives the position of 
 the central petal at $t=0$ :
 \begin{eqnarray} \label{suppl:alpha}
     \alpha(t=1/2,\tau,\Delta t)&=& \mathrm{tan}^{-1}\left(\frac{E_x(1/2)}{E_y(1/2)}\;\right)-\frac{\pi}{4} \; = \;\mathrm{tan}^{-1}\left(\frac{e^{-2\ln(2)\left(\frac{1/2-\tau/2}{\Delta t}\right)^2}}{e^{-2\ln(2)\left(\frac{1/2+\tau/2}{\Delta t}\right)^2}}\;\right)-\frac{\pi}{4}\label{eS1}.
 \end{eqnarray}
%
%
%

 It is worth mentioning that, for ultra-short pulses in the sub-two-cycle regime,
 the temporal position at $t=1/2$ opt. cycle is not well described 
by $T_0 \equiv 2\pi/\omega_0$ but by the \textit{principal period} $T_P \equiv 2\pi/\omega_P$, where $\omega_P$ refers to 
the \textit{principal frequency} 
 (see Refs.~\cite{neyra2021principal,PhysRevResearch.4.033254} for definition and in-depth analysis of these concepts). It is this period (frequency) that makes it possible to correctly locate the position of the maxima field amplitude for such ultra-short pulses. Also, it should be noted that the angle $\alpha$ results independent of $\tau$ when this is an integer multiple of $T_0$. Here, the AP pulse has right or left polarization (see Main Text for more details). 
\begin{figure}[h!]
\includegraphics[width=0.6\textwidth]{Fig2SM.eps}
\caption*{\textbf{Supplementary Fig. 2: Variation of the angle $\alpha$ as a function of the temporal delay $\tau$.} Red (blue) circles indicate the variation of the AP pulse for a temporal FWHM $\Delta t=2$ ($\Delta t=5$) opt. cycles. For each case, the triangles show the linear approximation. } \label{fig:Fig1Sa}
\end{figure}

From Eq.~\eqref{suppl:alpha} we can extract the linear approximation of $\alpha(\tau,\Delta t)$, that can be 
expressed as $\alpha_1(\tau,\Delta t)=\frac{ln(2)\tau}{\Delta t^2}$. 
In Supplementary Figure 2 we show the angle $\alpha$  as a function of the delay $\tau$ for each of the two AP pulses considered in this work. Red (blue) circles represent the case of FWHM $\Delta t=2$ ($\Delta t=5$) opt. cycles, while the corresponding linear approximation can be seen in red (blue) triangles. 
 As can be seen, $\alpha$ exhibits a linear-like behavior as a function of the delay $\tau$ in the case of $\Delta t=5$ opt. cycles. However, for $\Delta t=2$ opt. cycles the linearity holds, approximately, until $\tau\approx 2$ opt. cycles. 

\subsection{Time-resolved recollision angles}\label{suppl_angles}

From the results of the time-frequency analysis using the Gabor transform (see the Methods Section in the Main Text), here we calculate the time-resolved recollision angle of ionized electrons~\cite{murakami2013high}. This quantity can be derived as
\begin{equation}\label{SM_rec_angle}
\Theta(\omega,t) = \mathrm{tan}^{-1}\left(\sqrt{\frac{G_{y}(\omega,t)}{G_{x}(\omega,t)}}\;\right),
 \end{equation}
where we restrict ourselves to the first quadrant, i.e.~$\Theta(\omega,t) > 0$. This angle determines the angle at which the electron recombines with its parent ion. 

The complete wavelet analysis can be seen in Supplementary Figure 3, where we show the results for an AP pulse with a temporal FWHM of $\Delta t = 5$
opt. cycles and a global phase $\phi = 0$. In this figure, panels (a) and (b) depict the color maps of the Gabor transforms for the associated HHG spectrum, $G_{x}(\omega,t)$ and $G_{y}(\omega,t)$, while the color map in panel (c) indicates the corresponding value of $\Theta(\omega,t)$. It is worth mentioning that values of $\Theta(\omega,t)$ above, approximately, the 28th harmonic order, have no physical meaning since the emission probability is negligible (they are numerical artifacts that come from the computation of the expression in Eq.~\eqref{SM_rec_angle}). From panel (c), we can observe how is the variation of the emission angle between the different bursts of XUV radiation, which indicates the polarization angle of the different attosecond pulses in the APT. 
In agreement with what is seen in the Main Text, this angle goes from  
0$^\circ$ to 90$^\circ$, indicating the rotation of the polarization of the AP pulse in the $x-y$ plane. 
\begin{figure}[h!]
\includegraphics[width=0.8\textwidth]{Fig3SM_prueba.pdf}
    \caption*{\textbf{Supplementary Fig. 3: Time-frequency analysis of the twisted APT: multi-cycle pulses version.} Panels (a) and (b): wavelet analysis $G_x(\omega,t)$ and $G_y(\omega,t)$, obtained from the dipole accelerations $a_x(t)$ and $a_y(t)$, respectively, for an AP pulse with a temporal FWHM of $\Delta t = 5$ opt. cycles. The $x$ ($y$) component of the AP pulse is superimposed in a white (red) solid line. Panel (c): Recollision angles obtained from Eq.~(5). In this case, the color map scales from 0$^\circ$ to 90$^\circ$.}\label{fig:Fig3SMc}
\end{figure}

Supplementary Figure 4 shows the analysis analogous to the one performed in SM Figure 3, but for the case of an AP pulse with $\Delta t = 2$ opt. cycles. 
In panel (c), it can be seen the variation of the emission angle between the different bursts of XUV radiation, or attosecond pulses, which, as expected, is greater than in the previous case.  

Finally, in Supplementary Figure 5, we show the variation of the emission angle for a single burst which is generated, approximately, between 0 opt. cycles and 0.5 opt. cycles. From this figure, it can be observed the angle of the recombination for the ``short" and ``long" trajectories in a single burst, or in a half laser cycle. Between the 16th and the 20th harmonic order, the ``short" trajectories recombine with an angle of 40$^\circ$, while the ``long" trajectories do so at an angle of 50$^\circ$. Above the 20th harmonic order (cutoff region), the angle of recombination of both trajectories converges, approximately, to  45$^\circ$.

\begin{figure}[h!]
\includegraphics[width=0.8\textwidth]{Fig4SM_prueba.pdf}
\caption*{\textbf{Supplementary Fig. 4: Time-frequency analysis of the twisted APT: few-cycle pulses version.}  Same as Supplementary Figure 3 but for an AP pulse with a temporal FWHM of $\Delta t = 2$ opt. cycles. 
} \label{fig:Fig4SMd}
\end{figure}


\begin{figure}[h!]
\includegraphics[width=0.8\textwidth]{Fig5SM_new.png}
\caption*{\textbf{Supplementary Fig. 5: Time-frequency analysis of a single burst of the few-cycle pulse}. Recollision angles obtained from Eq. (5) for a single burst, approximately, between 0 and 0.5 opt. cycles, where the $x$ ($y$) component of the AP pulse (same AP pulse that was considered for the analysis in Supplementary Figure 4) is superimposed in white (red) solid line. The color map scale from 40$^\circ$ to 50$^\circ$.} \label{fig:Fig5SMd}
\end{figure}


\newpage

\bibliography{main}